\documentclass[12pt]{article}
\makeatletter \@addtoreset{equation}{section} \makeatother

\usepackage{epsfig,amsfonts}
\usepackage[fleqn]{amsmath}
\usepackage{amsthm,amssymb}
\usepackage{graphicx,psfrag}
\usepackage{hhline}
\usepackage{cite}

\def\be{\begin{equation}}
\def\ee{\end{equation}}
\def\bal{\begin{align}}
\def\eal{\end{align}}
\def\bea{\begin{eqnarray}}
\def\eea{\end{eqnarray}}

\newcommand {\zb} {{\bar z}}

\newcommand {\CalA} {\mathcal A}

\newcommand {\CalM} {\mathcal M}
\newcommand {\CalN} {\mathcal N}

\newcommand {\CalS} {\mathcal S}

\newcommand {\CalX} {\mathcal X}

\newcommand {\BC}   {\mathbb C}

\newcommand {\BR}   {\mathbb R}
\newcommand {\BP}   {\mathbb P}

\def\trace{\text{tr}}
\def\ri{{\rm i}}
\def\rd{{\rm d}}

\def\column#1#2{\left(\begin{array}[c]{cc}%
#1 &\\
#2 &
\end{array}\!\!\!\!\!\!\right)}

\def\matrix#1#2#3#4{\left(\begin{array}[c]{cc}%
#1 & #2\\
#3 & #4
\end{array}\right)}

\def\zb{\bar z}

\begin{document}

\begin{titlepage}
\begin{flushright}
RUNHETC-2013-15\\
\end{flushright}

\vspace{1.cm}

\begin{center}
\begin{LARGE}
{\bf Classical Conformal Blocks and Painlev\'e VI}

\end{LARGE}

\vspace{1.3cm}

\begin{large}

\textbf{Alexey Litvinov$^{1}$, Sergei Lukyanov$^{1}$,\\
Nikita
Nekrasov$^{2,\,3^{*},\,4^{*},\,5^{*}}$ and Alexander
Zamolodchikov$^{1^{*},\,3}$}

\end{large}

\vspace{1.cm}

$\ {}^{1}$NHETC, Department of Physics and Astronomy, Rutgers University,\\
     Piscataway, NJ 08855-0849, USA\\

${}^{2}$ Simons Center for Geometry and Physics,
Stony Brook, NY 11794-3636, USA\\

${}^{3}$  Institut des Hautes Etudes Scientifiques,  Bures-sur-Yvette 91440, France\\

${}^{4}$ Kharkevich Institute for Information Transmission Problems, Lab. 5, Moscow 127994 Russia\\

$^{5}$ Alikhanov Institute of Theoretical and Experimental Physics,  Moscow 117218, Russia\\

\vspace{.8cm}

\end{center}

\begin{center}
\centerline{\bf Abstract} \vspace{.8cm}
\parbox{13cm}{
We study the classical $c\to \infty$ limit of the Virasoro conformal blocks.
We point out that the classical limit of the simplest nontrivial null-vector
decoupling equation on a sphere leads to the Painlev\'e VI equation. This
gives the explicit representation of generic four-point classical conformal
block in terms of the regularized action evaluated on certain solution of the
Painlev\'e VI equation. As a simple consequence, the monodromy problem of the
Heun equation is related to the connection problem for the Painlev\'e VI.}
\end{center}

\vspace{.8cm}


\vfill
\begin{flushleft}
\rule{3.1 in}{.007 in}\\
$^{*}$ \text{On leave of absence}
\end{flushleft}

\end{titlepage}
\newpage

\section{\label{sec1}Introduction}

Originally, conformal blocks were introduced in the context of two
dimensional conformal field theories \cite{Belavin:1984vu}, where they play
fundamental role in the holomorphic factorization of the correlation
functions. More recently, these functions attracted renewed attention because
of their remarkable relation to the correlators of supersymmetry-protected
chiral operators in the ${\CalN}=2$ supersymmetric gauge theories in four
dimensions. This relation, dubbed the BPS/CFT correspondence in
\cite{Nekrasov:2004sem}, following the prior work in \cite{Nekrasov:2002qd,
Losev:2003py, Nekrasov:2003rj}, became a subject of intense development after
the seminal work \cite{Alday:2009aq} where the instanton partition functions
of the $\CalS$-class $\CalN=2$ gauge theories in the $\Omega$-background
\cite{Nekrasov:2002qd} were conjectured to be the Liouville (and, more
generally, the $ADE$ Toda) conformal blocks.

{}The conformal blocks are fully determined  by their defining properties
(the conformal symmetry in CFT, and, for the quiver gauge theories, the
instanton integrals of the supersymmetric gauge theories). However, apart
from (numerous) special cases, no closed form expressions are known beyond
the power series expansions. Therefore, any relations which could provide an
additional analytic control are of interest. In this note we establish a
relation between the so-called classical conformal blocks
\cite{Zamolodchikov:1987, Zamolodchikov:1995aa} and the classical action
evaluated on the special solutions of the celebrated Painlev\'e VI
equation.\footnote{This is not the first time the Painlev\'e VI emerges in
connection with conformal blocks. In \cite{Gamayun:2012ma} the associated
tau-function is shown to generate the conformal blocks of the $c=1$ CFT. This
is very interesting yet different from the relation between the classical
limit ($c\to \infty$) of the conformal blocks and classical action of the
Painlev\'e VI which we discuss here. The emergence of Painlev\'e VI in
connection to the classical conformal blocks was also pointed out in recent
paper \cite{kashani-poor}.}

\begin{figure}
[!ht]
\centering
\psfrag{z1}{$z_{1},\Delta_{1}$}
\psfrag{z2}{$z_{2},\Delta_{2}$}
\psfrag{z3}{$z_{n-1},\Delta_{n-1}$}
\psfrag{z4}{$z_{n},\Delta_{n}$}
\psfrag{p1}{$\Delta(P_{1})$}
\psfrag{p2}{$\Delta(P_{2})$}
\psfrag{p3}{$\Delta(P_{n-3})$}
\includegraphics[width=14  cm]{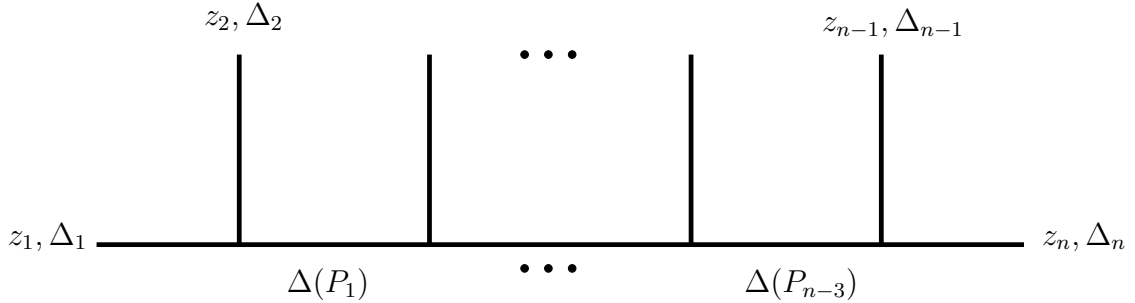}
\caption{\small Dual diagram representing the ``pant decomposition''
of $n$-punctured sphere for the
conformal block \eqref{vcorrpi}. The external legs are
associated with the insertions $V_{\Delta_i}(z_i),\ i=1,...,n$,
while the internal links represent the intermediate spaces with the primary dimensions
$\Delta(P_\alpha)$.
The vertices correspond to
the elementary three-punctured spheres of the given pant decomposition.}
\label{fig1}
\end{figure}

\begin{figure}
\centering
\psfrag{z1}{$0,\Delta_{1}$}
\psfrag{z2}{$x,\Delta_{2}$}
\psfrag{z3}{$\Delta(P)$}
\psfrag{z4}{$1,\Delta_{3}$}
\psfrag{z5}{$\infty,\Delta_{4}$}
\includegraphics[width=10  cm]{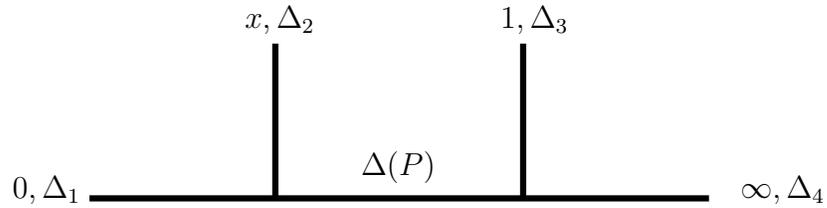}
\caption{\small The dual diagram representation of the four-point
conformal block $\mathcal{F}_P(x)$.}
\label{fig2}
\end{figure}

{}Generally, the conformal blocks are associated with
the moduli spaces ${\CalM}_{g,n}$ of
genus $g$ Riemann surfaces with $n$ punctures.
In the present note  we limit our
attention to the $n$-punctured spheres (and ultimately our result will be concerned
with the simplest nontrivial case $n=4$). Loosely speaking, the conformal
blocks are $n$-point correlation functions $\langle\,V_{\Delta_1}(z_1) \cdots
V_{\Delta_n}(z_n)\,\rangle$ of the chiral primary operators $V_\Delta(z)$.
The subscripts $\Delta$ indicate the associated conformal dimensions. Since
generally the chiral primaries are not local fields, the notion of their
correlation functions is ambiguous. Actual conformal blocks are defined
relative to a given ``pant decomposition'' of the $n$-punctured sphere,
usually represented by a ``dual diagram''. For example, the diagram in
Fig.\,\ref{fig1}
represents
the instruction to include only the states from the irreducible
representations with the conformal weights $\Delta(P_\alpha)$ in the
intermediate-state decompositions of the operator product,
\begin{eqnarray}\label{vcorrpi}
\mathcal{F}_{\bf P}({\bf z}) =\langle\,V_{\Delta_1}(z_1)\Pi_{\Delta(P_1)}
 V_{\Delta_2}(z_2)\Pi_{\Delta(P_2)} \cdots \Pi_{\Delta(P_{n-3})}
 V_{\Delta_n}(z_n)\,\rangle\,,
\end{eqnarray}
where $\Pi_{\Delta(P)}$ stand for the projection operators. (In what follows we
often omit the projection operators and just refer to the associated dual
diagram.) Here
and below we use the Liouville-inspired
parameterization of the ``intermediate'' dimensions
$\Delta(P)$ and the Virasoro central charge $c$,
\begin{eqnarray}\label{DeltaP}
\Delta(P) = \frac{c-1}{24} +P^2\,, \qquad c = 1+6\,(b+b^{-1})^2\,.
\end{eqnarray}
Thus, the $n$-point conformal block depends on $n-3$ parameters ${\bf P} = \{
P_1, P_2, \ldots P_{n-3}\}$. Although the definition \eqref{vcorrpi} involves
$n$ points $z_1,\ldots z_n$, projective transformations allow one to fix
three of them, e.g. by sending the three points, say $z_{n-2},\ z_{n-1},\
z_n$, to the standard locations $0,\ 1$ and $\infty$. Therefore, the
conformal block \eqref{vcorrpi} depends on $n-3$ moduli, which we
collectively denote by ${\bf z}=\{z_1, \ldots  z_{n-3}\}$. Although the
conformal block also depends on $n$ ``external leg'' dimensions $\Delta_1,
\Delta_2, \ldots \Delta_n$, we omit these parameters in the above notation
$\mathcal{F}_{\bf P}({\bf z}) $. In the simplest nontrivial case $n=4$ there
is only one complex modulus. In this special case we fix the coordinate $x$
on the moduli space ${\CalM}_{0,4}$ by setting
\begin{eqnarray}\label{z1..4}
\{z_1,\, z_2,\, z_3,\, z_4\} = \{ 0, \,x, \,1,\, \infty\}\,.
\end{eqnarray}
Also, there is only one intermediate state parameter, the momentum $P$.
Suppressing again the dependence on $\Delta_i,\ i=1, \ldots, 4$, we denote by
$\mathcal{F}_P (x)$ the $4$-point conformal block associated with the dual
diagram in Fig.\,\ref{fig2}.
The conformal properties
allow one to determine, in principle, any coefficient in the power series
expansion in the moduli, e.g. in
\begin{eqnarray}\label{cblock4}
\mathcal{F}_P (x) = x^{\Delta(P)-\Delta_1 - \Delta_2} \,\sum_{n=0}^\infty
\,\mathcal{F}_{P}^{(n)}\,x^n\,.
\end{eqnarray}
The recursion of \cite{Zamolodchikov:1985ie} allows one to generate the
coefficients in a fast and efficient way, especially numerically. The AGT
 conjecture \cite{Alday:2009aq} provides another powerful combinatorial
representation for these coefficients, using the explicit expression for
the instanton partition sums in the supersymmetric gauge theories
\cite{Nekrasov:2002qd} found using the fixed point methods.
The equivalence of the two representations for the linear quiver theories was
recently proven in \cite{Alba:2011}.

The classical limits of the conformal blocks appear when the Virasoro central
charge $c$ goes to infinity along with all the dimensions, so that the ratios
$\Delta_i/c$ and $\Delta(P_\alpha)/c$ remain fixed. In what follows it will be
convenient to regard $b^2$ defined in \eqref{DeltaP} as the Planck's constant.
Thus, the classical limit corresponds to $b \to 0$ and
\begin{eqnarray}\label{deltanu}
\Delta_i \to {\textstyle \frac{1}{b^2}}\ \delta_i\,, \qquad \Delta(P_\alpha) \to
{\textstyle \frac{1}{b^2}}\ \delta_{\nu_\alpha}
\end{eqnarray}
with finite ``classical dimensions'' $\delta_i$ and $\delta_\nu=
\frac{1-\nu^2}{4}$. The new classical parameter $\nu$ relates to $P$ as $P^2
= -\nu^2/4b^2$ (In this discussion the parameters $P$ and $\nu$ are generally
regarded as complex numbers). In the classical limit thus defined the
$n$-point conformal blocks are expected to exponentiate as\footnote{ The
exponentiation \eqref{Ftof} was conjectured in \cite{Zamolodchikov:1985ie,
Zamolodchikov:1995aa} and supported by the analysis of a long power series in $x$
generated via the recurrent relation of
\cite{Zamolodchikov:1985ie,Zamolodchikov:1987}. This exponentiation is
essential for the representation of the classical Liouville action in terms
of the classical conformal blocks. New support comes from the AGT example of
the BPS/CFT correspondence. On the gauge theory side the combinatorial
representation for the power series coefficients has the form of the virial
expansion for a gas of one-dimensional particles with a short-range
interaction. From that point of view the exponentiation \eqref{Ftof} is the
statement of the existence of the thermodynamic limit. This idea leads to the
equation for the partition function similar to the thermodynamic Bethe ansatz
equation \cite{Nekrasov:2009rc}. Recently another proof was found in
\cite{Nekrasov:2013} using the generalization of the limit shape equations.}
\begin{eqnarray}\label{Ftof}
\mathcal{F}_{\bf P}({\bf z})  \ \sim  \ \exp\big(\,{\textstyle \frac{1}{b^2}}\,\,
f_{{\boldsymbol \nu}}({\bf z})\,\big)\, ,
\end{eqnarray}
where the ``classical conformal block'' $f_{{\boldsymbol \nu}}({\bf z})$ depends on the
moduli ${\bf z}$, the parameters ${\boldsymbol \nu} = \{\nu_1, \nu_2, ... \nu_{n-3}\}$,
and $n$ external ``classical dimensions'' $\delta_i$.

Classical conformal blocks are of interest from several points of view. They
offer solution to the monodromy problem for the linear second order
differential equations with $n$ regular singularities, as we explain below.
The solution of classical Liouville equation and related uniformization
problem can be found in terms of the classical conformal blocks via certain
Legendre transform \cite{Zamolodchikov:1995aa}.
On the gauge theory side of the AGT correspondence, the
classical conformal blocks, or rather closely related functions
\begin{eqnarray}\label{wdef}
W({\boldsymbol \nu},{\bf z}) = W_{0}({\boldsymbol \nu}) +f_{{\boldsymbol \nu}}({\bf z})\ ,
\end{eqnarray}
where $W_{0}({\boldsymbol \nu})$ is
related to the $b \to 0$ limit of the $3$-point structure functions, it
can also be interpreted as
the perturbative contribution to the twisted
superpotential on the $\mathbb{SU}(2)$ gauge
theory side. Classical conformal blocks (or rather the twisted
superpotentials) can be interpreted in the
language of the symplectic geometry of the moduli
space of $\mathbb{SL}(2)$-flat connections, as explained in \cite{Nekrasov:2011bc}.

In the simplest case $n=4$ the classical conformal block $f_\nu (x)$ depends
on a single intermediate classical dimension $\delta_\nu =
\frac{1-\nu^2}{4}$, and a single modulus $x$ (both will be generally treated
here as complex numbers). With the conventional normalization
$\mathcal{F}_P^{(0)}=1$ of the conformal block \eqref{cblock4}, the classical
conformal block defined by \eqref{Ftof} behaves as
\begin{eqnarray}\label{fass}
f_\nu(x) = (\delta_\nu-\delta_1-\delta_2)\,\log x + O(x) \qquad \text {as}\quad x\to 0\,.
\end{eqnarray}
The main goal of this note is to show that this function is given by
(regularized) classical action evaluated on certain solution of the
Painlev\'e VI equation, which we specify in Section\,\ref{sec3}.

\section{\label{sect2}Classical Conformal Blocks and Monodromies of Ordinary Differential
Equations}

The classical conformal blocks are closely related to the monodromy problem
for ordinary linear differential equations. Consider the second order
differential equation with $n$ regular singularities,
\begin{eqnarray}\label{diff0}
\psi'' (z) + t(z) \psi(z) =0\,, \qquad t(z)=\sum_{i=1}^n\,\bigg(\,
\frac{\delta_i}{(z-z_i)^2}+\frac{c_i}{z-z_i}\,\bigg)\ .
\end{eqnarray}
The variable $z$ can be regarded as the complex coordinate on
$\mathbb{CP}^{1} \,\backslash\{z_1,z_2,\ldots z_n\}$, the Riemann sphere with $n$
punctures. The parameters $\delta_i$ will be identified with the classical
dimensions in \eqref{diff0}, and in this discussion we will always regard
them as fixed numbers. The coefficients $c_i$ are often referred to as the
``accessory parameters''; they, along with the positions of the singularities
$z_i$, are treated as the variables. The accessory parameters $c_i$ are
constrained by three elementary relations
\begin{equation}
 \sum_{i=1}^{n} c_{i} = 0\,, \qquad \sum_{i=1}^{n}( c_i\,z_{i} + {\delta}_{i}) = 0 \,,
 \qquad \sum_{i=1}^{n}( c_i\, z_{i}^{2} + 2\,{\delta}_{i}\,z_{i}) = 0\,,
\label{eq:sltwomom}
\end{equation}
 ensuring that $t(z)$ has no
additional singularity at $z=\infty$. Thus only $n-3$ of these parameters,
say $c_1, \ldots c_{n-3}$, are independent. Also, the projective
transformations of the variable $z$ in \eqref{diff0} allow one to send three
of the points $z_i$, say $z_{n-2}, z_{n-1}, z_n$, to the predesigned
positions, usually $0,1,\infty$. Therefore, with $\delta_i$ fixed, the
differential equation essentially depends on $2\,(n-3)$ complex parameters
$c_i,\, z_i$, $\ i=1,2,\ldots n-3$.

The differential equation \eqref{diff0} generates the ``monodromy group'' -- the
homomorphism of the fundamental group
\begin{eqnarray}\label{mdef}
M :\ \
{\pi}_{1}(\mathbb{CP}^{1}
\backslash\{z_i\}) \longrightarrow  \mathbb{ SL}(2,\mathbb{ C}).
\end{eqnarray}
To define the homomorphism precisely one has to pick a point $z_{*} \in
{\BC\BP}^{1}$, distinct from $z_{1}, \ldots  z_{n}$. If
$\Psi(z_{*})=(\psi_1(z_{*}), \psi_2(z_{*}))$ is some basis in the vector
space of solutions of \eqref{diff0}, specified by the local data at the point
$z_{*}$, then the continuation along any closed path defines the monodromy
matrix $M(\gamma)$: $\Psi(\gamma\circ z_{*}) = \Psi(z_{*}) M(\gamma)$, which
depends only on the homotopy class $\gamma \in \pi_1
(\mathbb{CP}^{1}\,\backslash\{z_i\})$ (the fundamental group defined relative
to the marked point $z_{*}$)  of the path. Let $\gamma_i \in \pi_1
(\mathbb{CP}^{1}\backslash\{z_i\}),\ i=1,2,\ldots n$ be the elementary paths
around the points $z_i$, and $M(\gamma_i) \in \mathbb{ SL}(2,\mathbb{ C})$
the associated elements of the monodromy group of \eqref{diff0}. If we change
the marked point $z_{*}$ to another point $z_{*}^{\prime}$, then the matrices
$M({\gamma}_{i})$ change as well. However, the change is the simultaneous
conjugation of all $M({\gamma}_{i})$ by the same $ \mathbb{ SL}(2,\mathbb{
C})$ element $h$ (which is the holonomy of \eqref{diff0} from $z_{*}$ to
$z_{*}^{\prime}$), $ M({\gamma}_{i}) \mapsto h M({\gamma}_{i}) h^{-1} $. The
parameters $\delta_i = (1-\lambda_{i}^2)/4$ determine the conjugacy classes
of $M_i$ via the equation
\begin{eqnarray}
m_{i} = {\rm Tr} \big(M(\gamma_i)\big) =
- 2\cos(\pi\lambda_i) \ .
\label{eq:trmi}
\end{eqnarray} With these parameters fixed, the space of such
homomorphisms, taken modulo overall conjugation, is essentially isomorphic to
 the moduli space of flat $\mathbb{ SL}(2,\mathbb{ C})$
connections $\mathcal{A}_{0,n}$ \cite{ab}
 \footnote{The moduli space ${\CalA}_{0,n}$ of
flat connections is the space of all $\mathbb{ SL}(2,\mathbb{ C})$ gauge
fields $A$ on the $n$-punctured sphere, having vanishing curvature $F_{A} =
\rd A + A \wedge A = 0$ and considered up to the gauge transformations $A
\mapsto g^{-1} A g + g^{-1} \rd g$, with $g (z, {\zb}) \in \mathbb{
SL}(2,\mathbb{ C})$. With proper restrictions on the behavior of $A$ and $g$
near the punctures, the moduli space $\CalA$ is isomorphic to the so-called
character variety, the space of all homomorphisms \eqref{mdef} considered up
to the conjugation: $M \sim h^{-1} M h$ for some $h \in  \mathbb{
SL}(2,\mathbb{ C})$.}. The point of this space can be parameterized by the
invariants ${\rm Tr}\big(M(\gamma_{a_1}\ldots \gamma_{a_N})\big)$, which obey
certain polynomial relations. It is well known that for the $n$-punctured
sphere the complex dimension of this space is exactly $2\,(n-3)$. This means
that the differential equation \eqref{diff0} generally does not admit
continuous isomonodromic deformations, and $2\,(n-3)$ parameters $c_i, z_i$
$(i=1,2,\ldots n-3)$ in \eqref{diff0} can be taken as local coordinates on at
least a part of the moduli space ${\CalA}$ of the flat connections.

The moduli space $\CalA_{0,n}$ admits a natural symplectic form, due to
Atiyah and Bott \cite{ab}. It turns out that
the parameters $c_i, z_i$ are canonically conjugate, i.e.
Darboux coordinates with respect to this form
\cite{Bilal:1990wn,boalch,Nekrasov:2011bc,Krichever:2001cx}
\begin{eqnarray}
\Omega = \sum_{i=1}^{n-3}\,\rd c_i\wedge \rd z_i\,.
\end{eqnarray}

To connect the classical conformal block in \eqref{Ftof} to the differential
equation \eqref{diff0}, recall that in general case of $c$ and at special
values of $\Delta$ the chiral vertex operators $V_\Delta (z)$ correspond to
the highest weight vectors of the so called ``degenerate representations'' of
the Virasoro algebra. In such cases the Verma module contains null vectors,
and as the result the conformal blocks involving such degenerate vertex
operators satisfy special differential equations. The simplest nontrivial
cases correspond to the null vectors on the level 2, which appear at two
values of $\Delta$, $\Delta_{(1,2)}$ and $\Delta_{(2,1)}$,
\begin{eqnarray}
&&\Delta_{(1,2)} = -\frac{1}{2}-\frac{3b^2}{4}\,, \qquad
\mid \text{null}\,\rangle = \left(b^{-2}\,L_{-1}^2 +L_{-2}\right)
\mid\Delta_{(1,2)}\,\rangle\,,\nonumber\\
&&\Delta_{(2,1)} = -\frac{1}{2}-\frac{3}{4b^2}\,, \qquad
|\, \text{null}\,\rangle = \left(b^2\,\, L_{-1}^2 +L_{-2}\right)
|\, \Delta_{(2,1)}\,\rangle\,,
\end{eqnarray}
where we also display the associated null-vectors. The null-vector
decoupling leads to the second-order differential equations for the conformal
blocks, e.g.
\begin{eqnarray}\label{pde12}
\left[\,\frac{1}{b^2}\,\frac{\partial^2}{\partial z^2} + \sum_{i=1}^n
\left(\,\frac{\Delta_i}{(z-z_i)^2}+\frac{1}{z-z_i}\,\frac{\partial}{\partial z_i}
\,\right)\,\right]\, \langle\,
V_{(1,2)}(z) V_{\Delta_1}(z_1) \ldots  V_{\Delta_n}(z_n)\,\rangle =0\,,
\end{eqnarray}
where $V_{(1,2)} = V_{\Delta_{(1,2)}}$. Conformal blocks with the vertex
$V_{(2,1)}$ obey similar differential equation with $b^2$ replaced by
$b^{-2}$. The differential equation itself does not depend on the ``dual
diagram'' which must be added to specify the conformal block; the later comes
through the choice of the solution. Note that unlike $\Delta_i$, the
dimension $\Delta_{(1,2)}$ approaches finite limit $-1/2$ as $b \to 0$. The
usual semiclassical intuition then suggests the following semiclassical form
of the $n+1\,$-point conformal block involving $V_{(1,2)}$,
\begin{eqnarray}\label{psicorr}
\langle\,V_{(1,2)}(z) V_{\Delta_1}(z_1)\ldots V_{\Delta_n}(z_n)\,\rangle_{b\to 0}
\to \psi (z|\, {\bf z})\,
\exp\big(\,{\textstyle\frac{1}{b^2}}\ \,f_{\boldsymbol \nu}({\bf z})\,\big)\,,
\end{eqnarray}
where $f_{\boldsymbol \nu}({\bf z})$ is the same classical conformal block as in
\eqref{Ftof}. Then the consistency with the null-vector decoupling equation
\eqref{pde12} requires that the accessory parameters are determined in terms
of the classical conformal block as follows
\begin{eqnarray}\label{gradf}
c_i = \frac{\partial}{\partial z_i}\, f_{\boldsymbol \nu}({\bf z})\,.
\end{eqnarray}
Eq.\eqref{gradf} is similar to the equation appearing in the context of
the uniformization problem \cite{zt}, but the associated monodromy problem
for \eqref{diff0} is quite different; in particular the ``action'' $f_{\boldsymbol \nu}({\bf
z})$ has different meaning. In the classic uniformization problem
\cite{klein,poincare} one is interested in the set $\bf c$ of accessory
parameters $\{ c_i\}$ such that the monodromy group of \eqref{diff0} is Fuchsian
(which means it can be embedded into a real subgroup of $\mathbb{SU}(2)$ or
$\mathbb{SL}(2,\mathbb{R})$). This monodromy problem involves the reality
condition, and the corresponding accessory parameters obey \eqref{diff0} with
the real action $f({\bf z}, {\overline{\bf z}})$. The latter coincides with
the Liouville action on the $n$-punctured sphere \cite{zt}. In our case
$f_{\boldsymbol \nu}({\bf z})$ is the classical conformal block, and the
accessory parameters given by \eqref{gradf} solve another monodromy problem,
which is holomorphic in ${\bf z}$, and involves $n-3$ additional parameters
${\boldsymbol \nu}$, as follows.\footnote{The accessory parameters associated
with the uniformization problem can also be expressed in terms of the
classical conformal blocks through the Legendre transform, as is explained in
\cite{Zamolodchikov:1995aa}. This Legendre transform can be understood as
describing the ${\mathbb SL}(2, {\BR})$ connections in the coordinates of
\cite{Nekrasov:2011bc} as the real slice.}

\begin{figure}
\centering
\includegraphics[width=10  cm]{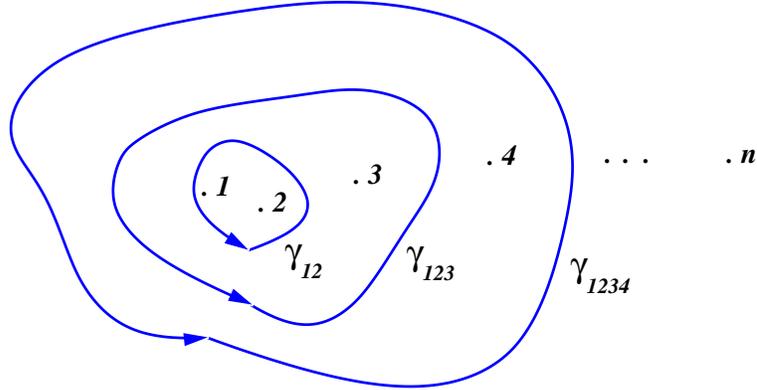}
\caption{\small The elements $\gamma_{12},\ \gamma_{123}, \ldots  \gamma_{12...n-2}$ of
the fundamental group ${\pi}_1(\mathbb{CP}^{1} \backslash\{z_i\})$. Choosing the accessory
parameters in \eqref{diff0} according to \eqref{gradf} fixes the conjugacy classes of the
associated elements of the monodromy group of \eqref{diff0} as given in \eqref{cgammas}.}
\label{fig3}
\end{figure}

The choice of the accessory parameters according to \eqref{gradf} locally
defines $n-3$ (complex) dimensional subspace in the $2\,(n-3)$ dimensional
moduli space of flat connections. Recall that the definition of the conformal
block $\mathcal{F}_{\bf P}({\bf z})$, and ultimately of its classical limit
$f_{\boldsymbol \nu}({\bf z})$, involves the ``dual diagram''. In this discussion we
assume the ``haircomb'' diagram in Fig.\,\ref{fig1}, so that the parameters
${\boldsymbol \nu} = \{\nu_1, \nu_2, \ldots \nu_{n-3}\}$ are inherited, in the classical limit,
from the dimensions $\{\Delta(P_1), \ldots  \Delta(P_{n-3})\}$ in the
intermediate state decomposition \eqref{vcorrpi}. Therefore the parameters
$\boldsymbol \nu$ have a simple interpretation in terms of the
monodromy of the the $n+1\,$-point conformal block \eqref{psicorr} under the
continuations in the variable $z$. Namely, let $M_{12}=M(\gamma_{12})$,
$M_{123} = M(\gamma_{123})$, \ldots  $M_{123\ldots n-2}=M(\gamma_{12\ldots
n-2})$ be the the monodromy matrices associated with the paths
$\gamma_{12\ldots k} = \gamma_1\circ \gamma_2 \circ \ldots \circ \gamma_k$
shown in Fig.\,\ref{fig3}.
The choice \eqref{gradf} of the accessory parameters in
\eqref{diff0} fixes the conjugacy classes\footnote{This follows from the well
known ``fusion rules'' for the operator product expansions involving the
degenerate operator $V_{(1,2)}$,
\begin{eqnarray}\nonumber
V_{(1,2)}(z)V_{\Delta(P)}(z_0) = (z-z_{0})^{b\alpha}\,V_{\Delta(P+\frac{\ri b}{2})}(z_0)
+(z-z_0)^{b(b+b^{-1}-\alpha)}\,
V_{\Delta(P-\frac{\ri b}{2})}(z_0)\,.
\end{eqnarray}
where $\alpha = \ri P+\frac 12 \left( b+b^{-1} \right)$.
In the classical limit $P\to -\ri\nu/2b$, so that the
exponents become $1/2\pm \nu/2$.}
\begin{eqnarray}\label{cgammas}
{\rm Tr}(M_{12}) = -2\,\cos(\pi\nu_1), \  {\rm Tr}(M_{123}) =
-2\,\cos(\pi\nu_2), \
 \ldots\   {\rm Tr}(M_{12\ldots n-2}) = -2\,\cos(\pi\nu_{n-3})\,.
\end{eqnarray}

A neat geometric way to express the above statement is as follows
\cite{Nekrasov:2011bc}. The $n-3$ monodromy parameters $\nu_1, \ldots
\nu_{n-3}$ can be taken as a half of local coordinates on the moduli space of
flat connections. It is easy to see that these variables are
Poisson-commuting with respect to the Atiyah-Bott symplectic
form.\footnote{This follows from the ultralocal form of the Atiyah-Bott
symplectic form $\Omega = \frac{1}{2\pi\ri}\ \int \ {\trace}(\delta
A \wedge  \delta A )$, since one can always choose non-intersecting
paths $\gamma_{12\ldots k}$, as in Fig.\,\ref{fig3}.}

One can then define, using a
geometric construction
involving an $n$-gon in the group $\mathbb{SL}(2,{\BC})$
 \cite{Nekrasov:2011bc}, a set of
canonically conjugate variables
 ${\mu}_{1}, \ldots  {\mu}_{n-3}$,\footnote{Note
that in notations of \cite{Nekrasov:2011bc} the coordinates ${\nu}_i$ are equal to
$1+{\alpha}_{i} / {\ri \pi}$ while ${\mu}_{i} = {\beta}_{i}/2$}
so that
\bea\label{iasusa}
\Omega =
\sum_{i=1}^{n-3}\,\rd\nu_i\wedge \rd\mu_i \ .
\eea
In view of
\eqref{gradf}, it is useful to introduce a
function \eqref{wdef}, with an appropriate choice of the term $W_0(\boldsymbol{\nu})$.
This function then generates
the canonical transformation between the coordinates $(c_i,z_i)$ to another
set of local coordinates $(\nu_i,\mu_i)$,
\begin{eqnarray}
\mu_i = \frac{\partial W}{\partial \nu_i}\,,
\qquad c_i =\frac{\partial W}{\partial z_i}\ .
\end{eqnarray}

\begin{figure}
\centering
\includegraphics[width=7  cm]{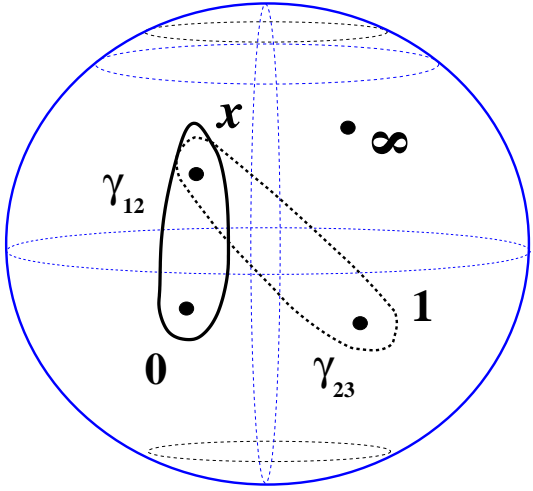}
\caption{\small Elements $\gamma_{12}=
\gamma_1\circ\gamma_2$ and $\gamma_{23}=\gamma_2\circ\gamma_3$
of the fundamental group of $\mathbb{CP}^1$ with
four punctures. The parameters $\nu,\ \mu$
 describe the conjugacy classes of
$M(\gamma_{12})$ and $M(\gamma_{23})$ via Eqs.\eqref{M12},
 \eqref{M23}.}
\label{fig4}
\end{figure}

In the simplest nontrivial case $n=4$ Eq.\eqref{diff0} is known as the Heun
equation,
\begin{eqnarray}\label{heun}
\psi''(z) + \left(\,\frac{\delta_1}{z^2}+\frac{\delta_2}{(z-x)^2}+\frac{\delta_3}
{(z-1)^2} + \frac{x(x-1)\,C}{z(z-1)(z-x)}-\frac{\delta_1+\delta_2 +
\delta_3-\delta_4}{z(z-1)}\, \right) \psi(z)=0\, .
\end{eqnarray}
The four-point classical conformal block $f_\nu(x)$ solves the monodromy
problem for this equation in the following sense. The choice
\begin{eqnarray}\label{cfx}
C = C(x,\nu)= \frac{\partial}{\partial x}\, f_\nu (x)
\end{eqnarray}
of the accessory parameter $C$ in \eqref{heun} fixes
the conjugacy class of the monodromy along the
path $\gamma_{12}=\gamma_1\circ\gamma_2$ (see Fig.\,\ref{fig4}),
\begin{eqnarray}\label{M12}
{\rm Tr}\big(M(\gamma_{12})\big) =-2\,\cos(\pi\nu)\,.
\end{eqnarray}
On the other hand
\begin{eqnarray}
{\mu} =  \frac{\partial}{\partial \nu}
\, \big( W_{0}({\nu}) + f_{\nu}(x)\big)
\label{eq:muco}
\end{eqnarray}
determines the conjugacy class of the monodromy along the path
$\gamma_{23}=\gamma_2\circ\gamma_3$ (see Fig.\,\ref{fig4}). With the choice
\bea\label{yoaisoaso}
W_0(\nu) =-\frac{1}{4}\ \int\rd\nu \ \log\big(\,\Omega(\nu,\lambda_1+\lambda_2)\,\Omega(\nu,\lambda_1-\lambda_2)
\,\Omega(\nu,\lambda_3+\lambda_4)\,\Omega(\nu,\lambda_3-\lambda_4)\,\big)\ ,
\eea
where
$$
\Omega(\nu,\lambda)=\frac{\Gamma(\frac{1}{2}+\frac{\nu}{2}+\frac{\lambda}{2})\,
\Gamma(\frac{1}{2}+\frac{\nu}{2}-\frac{\lambda}{2})}
{\Gamma(\frac{1}{2}-\frac{\nu}{2}+\frac{\lambda}{2})\,
\Gamma(\frac{1}{2}-\frac{\nu}{2}-\frac{\lambda}{2})}
\ \frac{\Gamma(1-\nu)}{\Gamma(1+\nu)}\ ,
$$
this relation reads, explicitly
(cf. \cite{Nekrasov:2011bc}):
\begin{eqnarray}\label{M23}
{\rm Tr}\big(M(\gamma_{23})\big) &= &-   \frac{\cosh(2\mu  )}{2\,{\sin}^2 ({\pi}{\nu})}\
\sqrt{c_{12}({\nu})c_{34}({\nu})}\\
&-&\frac{(m_{1}-m_{2})(m_{3}-m_{4})}{8\, {\cos}^{2}
(\frac{\pi\nu}{2})}+
\frac{(m_{1}+m_{2})(m_{3}+m_{4})}
{8\, {\sin}^{2} (\frac{\pi\nu}{2})}\ ,
\nonumber
\end{eqnarray}
where $m_i$'s were defined in \eqref{eq:trmi} and
\begin{equation}
c_{ij}({\nu}) =16\,
{\cos}\big( {\textstyle\frac{\pi}{2}}
({\nu} + {\lambda}_{i} + {\lambda}_{j} )\big)\, {\cos}
\big({\textstyle\frac{\pi}{2}}
({\nu} + {\lambda}_{i} - {\lambda}_{j} )\big)\,
{\cos}\big( {\textstyle\frac{\pi}{2}}
({\nu} - {\lambda}_{i} +
{\lambda}_{j} )\big)\,
{\cos} \big(
{\textstyle\frac{\pi}{2}}
({\nu} - {\lambda}_{i} - {\lambda}_{j} )\big)\ .\end{equation}
\section{\label{sec3}Hamilton-Jacobi equation and Painlev\'e VI}

\subsection{Classical limit of $V_{(2,1)}$}

The equation \eqref{diff0} has emerged in the classical limit from the
insertion of the degenerate vertex operator $V_{(1,2)}$. Inserting the other
level-2 degenerate operator $V_{(2,1)}$ has quite different effect. In the
limit $b\to 0$ its dimension $\Delta_{(2,1)}$ goes to infinity as $-3/4b^2$,
therefore the $n+1\,-\,$point conformal block with the $V_{(2,1)}$ insertion
exponentiates as
\begin{eqnarray}\label{wkbs}
\langle\,V_{(2,1)}(y) V_{\Delta_1}(z_1) \ldots V_{\Delta_n}(z_n)\,\rangle_{b\to 0}
\sim \exp\big(\,{\textstyle\frac{1}{b^2}}\, S(y,\{z_i\})\,\big)\,.
\end{eqnarray}
The null-vector decoupling equation has the form \eqref{pde12} but with $b$
replaced with $b^{-1}$, so that the coefficient in front of the second
derivative becomes large. This is fully consistent with the WKB form
\eqref{wkbs}, and leads to the Hamilton-Jacobi - like equation (see e.g. \cite{Teschner:2010je})
\begin{eqnarray}\label{hjn}
\left(\frac{\partial S}{\partial y}\right)^2 +\sum_{i=1}^n\,\left(\,\frac{\delta_i}
{(y-z_i)^2} + \frac{1}{y-z_i}\,\frac{\partial S}{\partial z_i}\,\right) =0\ .
\end{eqnarray}

Here we will show how this equation can be quite useful in the simplest
nontrivial case of $n=4$. Thus, we will start with the 5-point conformal
block with one of its insertions being $V_{(2,1)}(y)$. We use projective
transformations to move $z_1$, $z_3$, and $z_4$ to the positions $0$, $1$ and
$\infty$, respectively, and we denote $z_2=t$. In the classical limit we have
\begin{eqnarray}\label{5p21}
\langle \,V_{(2,1)}(y)V_{\Delta_1}(0)V_{\Delta_2}(t)
V_{\Delta_3}(1)V_{\Delta_{5}}(\infty)\,\rangle_{b\to 0} \ \sim \
\exp\big(\,{\textstyle\frac{1}{b^2}}\,{ S}(y,t)\,\big)\,.
\end{eqnarray}
At this point we do not specify the particular choice of the conformal block.
Note also that we denoted $\Delta_{5}$, not $\Delta_4$, the dimension
associated with $z_4=\infty$; the reason will become clear soon. Introducing
the function
\begin{eqnarray}\label{newaction}
&&\tilde{S}(y,t) =  { S}(y,t)+\varphi(y,t)\,,\\
&&\varphi(y,t)=
\big(\delta_1+\delta_2-{\textstyle\frac{1}{4}}\big)\,\log(t) +
\big({\textstyle\frac{1}{4}}+\delta_2+\delta_3-\delta_1-\delta_{5}\big)
\log(1-t)-{\textstyle\frac{1}{2}}\,\log\big(y(y-1)\big)\nonumber
\end{eqnarray}
allows one write the equation \eqref{hjn} as the conventional Hamilton-Jacobi
equation
\begin{eqnarray}
\frac{\partial {\tilde S}}{\partial t} + H\left(y,\frac{\partial {\tilde S}}{\partial y},t\right) =0
\end{eqnarray}
for a one-dimensional  dynamical system with the phase-space coordinates $(y,p)$ and
the time-dependent Hamiltonian
\begin{eqnarray}\label{Hp6}
H(y,p,t)=\frac{(y-t)y(1-y)}{t(1-t)}\ p^2-\frac{\delta_1-\frac{1}{4}}{y(1-t)}-
\frac{\delta_2}{t-y}-\frac{y(\delta_3-\frac{1}{4})}{t(y-1)}-
\frac{(\delta_{5}-\frac{1}{4})\, y}{t(1-t)}\ .
\end{eqnarray}
Here, as before, $\delta_i$ are classical dimensions associated with
$\Delta_i$ in \eqref{5p21}.

\subsection{Painlev\'e VI}

The Hamiltonian might not look terribly attractive, but the corresponding
equation of motion
\begin{eqnarray}\label{painleve6}
{\ddot y}&=&\frac{1}{2}\,\Big(\frac{1}{y}+\frac{1}{y-1}+\frac{1}{y-t}\Big)\, {\dot y}^2-
\Big(\frac{1}{t}+\frac{1}{t-1}+\frac{1}{y-t}\Big)\ {\dot y}\\
&+&\frac{2 y(y-1)(y-t)}{t^2(1-t)^2}\
\bigg(\,\big(\delta_1-{\textstyle \frac{1}{4}}\big)\,\frac{t}{y^2}
+\delta_2\,\frac{t(t-1)}{(y-t)^2}+
\big(\delta_3-{\textstyle \frac{1}{4}}\big)\,\frac{1-t}{(y-1)^2}+
{\textstyle \frac{1}{4}}-\delta_{5} \,\bigg)\nonumber
\end{eqnarray}
is recognized as the celebrated Painlev\'e VI equation. This is of course the
most general of the ordinary differential equations of the type ${\ddot y} =
R(y,{\dot y},t)$ whose solutions $y(t)$ have no ``movable singularities'' but
simple poles. Movable singularities are the singularities of a solution
$y(t)$, viewed as the function of complex $t$, whose locations depend on the
specific choice of the solution. Since by construction it is natural to regard the
variable $y$ as living on the Riemann sphere, the simple poles are not really
singularities, and one can say that the solutions of \eqref{painleve6} have
no movable singularities at all. Of course, there are ``fixed'' singularities
of the solutions $y(t)$ at $t=0,1,\infty$.

\subsection{Classical conformal block as Painlev\'e VI action}

\begin{figure}
\centering
\psfrag{z1}{\large{$0,\delta_{1}$}}
\psfrag{z2}{\large{$x,\delta_{2}$}}
\psfrag{z3}{\large{$\delta_{\nu}$}}
\psfrag{z4}{\large{$1,\delta_{3}$}}
\psfrag{z5}{\large{$\infty,\delta_{4}$}}
\includegraphics[width=10  cm]{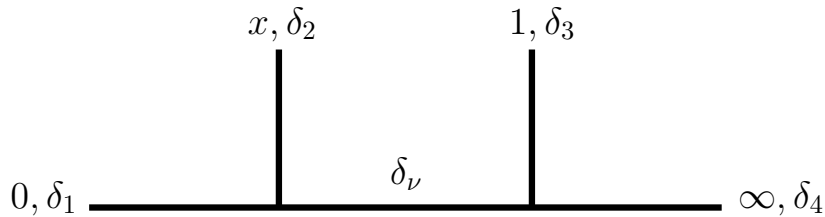}
\caption{\small Dual diagram representing the classical conformal block $f_\nu(x)$.}
\label{fig5}
\end{figure}

How this relation to the Painlev\'e VI can be useful in evaluating the
classical conformal block in Fig.\,\ref{fig5}?
Let us first make the following trivial observation. Let $y(t)$ be a
solution of \eqref{painleve6} such that
\begin{eqnarray}
y(t_1)=y_1\,, \qquad y(t_2) =y_2\,.
\end{eqnarray}
Then by the definition of the classical action  we have
\begin{eqnarray}
&&\langle\, V_{(2,1)}(y_2)\, V_{\Delta_1}(0)\,V_{\Delta_2}(t_2)\,
V_{\Delta_3}(1)\,V_{\Delta_5}(\infty)\,\rangle_{b\to 0}\ \sim\
\exp\Big[\, {\textstyle \frac{1}{b^2}}\,
\big(\,\varphi(y_1,t_1)-\varphi(y_2,t_2)\,\big)\,\Big]
\times\nonumber
\\
&&\ \ \ \ \exp\left(\, { \frac{1}{b^2}}\,
\int_{t_1}^{t_2}{\rm d} t\ L(y,{\dot y},t)\,\right)\ \ \
 \langle\, V_{(2,1)}(y_1)\, V_{\Delta_1}(0)\,V_{\Delta_2}(t_1)\,
 V_{\Delta_3}(1)\,V_{\Delta_5}(\infty)\,\rangle_{b\to 0}\ ,
\label{cblockact}
\end{eqnarray}
where $\varphi(y,t)$ is the same as in \eqref{newaction}, and $L(y,{\dot
y},t)$ is the Lagrangian associated with the Hamiltonian \eqref{Hp6}. As was
mentioned, general solution $y(t)$ is singular at $t=0,1,\infty$. Choose one
of these points, say $t=0$. Near this singularity $y(t)$ behaves as $y(t) \to
-\kappa\,t^{\nu}$ where $\kappa$ and $\nu$ are parameters of the solution.
The exponent $\nu$ always satisfies $0 \leq \Re  \,\nu \leq 1$,\footnote{At
$\nu=0, \,1$ the asymptotic involves logarithms. Here we ignore such
subtleties.} so that generally $y(t) \to 0$ as $t\to 0$. Therefore, if one
chooses $t_1=0$ in \eqref{cblockact}, three of the five points in the
conformal block in the r.h.s. collide, and thus it reduces to a constant. On
the other hand, the same trajectory $y(t)$ passes, at some nonsingular values
of $t = x \neq 0,1,\infty$, through the other insertion points $0$, $t$, $1$,
or $\infty$. At any such event two of the five insertions in the l.h.s.
five-point conformal block in \eqref{cblockact} merge, and it becomes a four
point block. Thus such trajectory $y(t)$ interpolates between the three-point
block (a constant) and a four point block which, according to
\eqref{cblockact} is then expressed essentially through the action
$\int_{0}^x\rd t\, L(y,{\dot y},t)$ evaluated on the trajectory. This outlines
our strategy.

Let us add some details. For a solution $y(t)$ we call the ``intercept
points'' the (generally complex) values of $t\neq 0,1,\infty$ at which $y(t)$
takes any of the values $y(t)=0,t,1,\infty$. Note that according to the
Painlev\'e property (no movable singularities) the intercept points are
regular point of $y(t)$ on the Riemann sphere. Out strategy requires setting
$t_2$ in \eqref{cblockact} equal to one of the intercept points. For
definiteness, assume that the intercept point closest to zero is a pole at
some $t=x$. The equation \eqref{painleve6} dictates two possibilities for the
residue at the pole, $y(t)\to \pm\,\frac {1}{\lambda_5}\,\frac{x(1-x)}{t-x}$
as $t\to x$, where $\lambda_5$ parameterizes the classical dimension
$\delta_5 = \frac{1-\lambda_{5}^2}{4}$. As explained above, at $t_2=x$ the
l.h.s. of \eqref{cblockact} becomes the four point conformal block
\begin{eqnarray}\label{4block}
\langle\,  V_{\Delta_1}(0)\,V_{\Delta_2}(x)\,
V_{\Delta_3}(1)\,V_{\Delta_4}(\infty)\,\rangle_{b\to 0}\ .
\end{eqnarray}
Here $\Delta_4$ is related to $\Delta_{5}$ in \eqref{cblockact} through the
well known fusion rule \cite{Belavin:1984vu}
\begin{eqnarray}\label{frule}
V_{(2,1)}V_{\Delta(P)} =\big[\,V_{\Delta(P+\ri/2b)}\,\big] + \big[\,V_{\Delta(P-\ri/2b)}\,\big]\ .
\end{eqnarray}
The fusion rule \eqref{frule} suggests two possible values for the classical
dimension $\delta_4$ associated with $\Delta_4$ in \eqref{4block}, namely
$\delta_4 = \frac{1- (\lambda_5\pm 1)^2}{4}$. In fact, in the classical limit
only one of the two terms in \eqref{frule} dominates. It turns out that at
all $\lambda_5$ with non-negative real part the dominating term corresponds
to $\lambda_{4} =\lambda_5+1$, at least for $x$ sufficiently close to $0$
(we have verified this statement numerically for real $\lambda_4$), so that
we have to take $\delta_5 =\frac{1-(\lambda_4-1)^2}{4}$ in order to
reproduce given $\delta_4 =\frac{1-\lambda_{4}^2}{4}$.

To summarize, in evaluating the classical conformal block emerging in the
limit $b\to 0$ of \eqref{4block}, we are interested in the solution of the
Painlev\'e VI equation \eqref{painleve6} with the following properties
\begin{eqnarray}\label{ydef}
&&y(t) \to -\kappa\,t^\nu \qquad\qquad \qquad  \qquad\
\ \ \  \ \ \ \ \    \text{as} \quad   t \to 0\,,\\
&&y(t) = \frac{1}{\lambda_4-1}\,\frac{x(1-x)}{t-x} + y_0 + \ldots  \qquad
\text{as} \quad t\to x \,,\label{ydefb}
\end{eqnarray}
where the dots in the second line represent a power series in $(t-x)$. In
\eqref{ydef} $\nu$ and $x$ should be regarded as independent parameters;
equation \eqref{painleve6} then determines $\kappa$ and $y_0$ as the
functions of these.\footnote{Unlike the ``initial data'' $(\nu,\kappa)$ or
$(x,y_0)$, the parameters $\nu,x$ do not fix the solution uniquely. There is
(likely infinite) discrete set of solutions with the same $\nu$ and $x$,
which correspond to different branches of the classical conformal block. We
hope to study this important question in the future.  In this work we always
consider $x$ sufficiently close to $0$, so that there is a unique ``natural''
choice of the solution.} Note that in \eqref{ydefb} we have chosen one of two
signs of the residue at $t=x$ consistent with \eqref{painleve6}; this choice
leads to the desired value $\frac{1-\lambda_{4}^2}{4}$ for the classical
dimension $\delta_4$, while the opposite sign of the residue would produce
$\frac{1-(2-\lambda_4)^2}{4}$. Once the solution is found, the classical
conformal block is expressed through the Painlev\'e VI action evaluated on
this solution, from $t=0$ to $t=x$, according to the Eq.\eqref{cblockact}.

\subsection{Regularized action and accessory parameter}

Although the solution $y(t)$ described in the previous subsection is singular
only at the singular points $0,1,\infty$, the Lagrangian $L(y,{\dot y},t)$
has poles at all intercept points. In particular, the action integral in
\eqref{cblockact}, when specified to $t_1=0$ and $t_2=x$, diverges
logarithmically at both ends of the integration contour. These divergences
are well expected. Thus, the divergence at $t\to x$ reproduces the classical
limit of the divergence of the operator product expansion
\begin{eqnarray}
V_{(2,1)}(y(t))V_{\Delta_5}(\infty)\ \sim\ (y(t))^{\Delta_4-\Delta_5-
\Delta_{(2,1)}}
\,V_{\Delta_4}(\infty) \qquad \text{as} \quad y(t) \to \infty
\end{eqnarray}
in \eqref{cblockact}, while the divergence at $t\to 0$ represents the
singular fusion of the three chiral vertices $V_{(2,1)}(y(t))
V_{\Delta_1}(0)V_{\Delta_2}(t)$.

This interpretation leads to the regularization prescription. One adds
$x$-independent counterterms to the classical action designed to compensate
for the divergences of the OPE coefficients. The result is the the following
regularized expression for the classical conformal block
\begin{eqnarray}\label{fact}
f_\nu(x)&=&(\delta_\nu-\delta_1-\delta_2)\ \log (x)+
(\delta_1-\delta_2-\delta_3+\delta_4)\ \log(1- x)\\
&+&
\frac{\nu}{2}\ \Big(\log\big(\kappa/\kappa_0\big)+\nu\,\log(x)\,\Big)+
\int_0^x\,d t\, \bigg(\,L(y,{\dot y},t)-\frac{\nu^2}{4t}-\frac{\lambda_4-1}
{2(x-t)}\,\bigg)\ ,
\nonumber
\end{eqnarray}
where $ \delta_i={\textstyle
\frac{1}{4}}\ (1-\lambda_i^2)$ and
\begin{eqnarray}
\kappa_0=\lim_{x\to 0}\big(x^\nu\,\kappa(x,\nu)\big)=\frac{4\, \nu^2}{(\nu-1-\lambda_3+
\lambda_4)(\nu-1+\lambda_3+\lambda_4)}
\end{eqnarray}
is fixed by the condition \eqref{fass} (see Eqs.\,\eqref{u0},\,\eqref{Akappa}
and \eqref{xnuA} below).

This formula looks somewhat difficult for practical applications. However, if
one is interested in the accessory parameter \eqref{cfx}, it is possible to
bypass altogether the evaluation of the action integral in \eqref{fact}.
Indeed, the derivative of the action with respect to $x$ generates two terms.
The first, easy term comes from the upper integration limit, while the
second, more difficult terms accounts for the derivative of the solution
$y(t)$ itself with respect to $x$. But since the action is stationary with
respect to the variations around the trajectory $y(t)$, this second term
reduces to the boundary contribution. As the result, one finds
\begin{eqnarray}\label{cxnu}
C(x,\nu) = \frac{\partial f_\nu(x)}{\partial x}= \frac{(\lambda_4-1)^2}{x (1-x)}\ (x-y_0) +
\frac{\delta_3-\delta_1+\delta_2-\delta_4}{x}+
\frac{\delta_2+\delta_3+\delta_4-\delta_1}{1-x}\ ,
\end{eqnarray}
where $y_0=y_0(x,\nu)$ is the constant term in the expansion \eqref{ydefb}.
Thus, finding the accessory parameter $C(x,\nu)$ is closely related to
solving the connection problem for the Painlev\'e VI equation, i.e. relating
the parameters $(x,y_0)$ to $(\nu,\kappa)$. Similar calculation yields
\begin{eqnarray}\label{muxnu}
\frac{\partial f_\nu (x)}{\partial \nu} =\frac{1}{2}\ \log(\kappa)-\frac{1}{2}\ \frac{\partial}{\partial\nu}\, \big(\nu\,\log(\kappa_0)\big)\ .
\end{eqnarray}
We note that Eqs.\eqref{cxnu}, \eqref{muxnu} generalize the equations (4.36)
of Ref.\cite{Lukyanov2011} obtained previously for the monodromy of the
Mathieu equation and connection problem of Painlev\'e III.

\subsection{Power series}

As is well known  (see e.g.\cite{Guzzetti2011}),
in the vicinity of the singular point, e.g. near
$t=0$, the solution $y(t)$ with the asymptotic behavior \eqref{ydef} expands
in a double power series
\begin{eqnarray}\label{doubleser}
\frac{1}{y(t)}=\sum_{n=0}^\infty t^n\ U_n(t)\,, \qquad
U_n(t)=\sum_{m=-n-1}^{n+1}t^{-\nu m}\ U_{n,m}\,,
\end{eqnarray}
where all the coefficients $U_{n,m}$ are uniquely determined by the equation
\eqref{painleve6} once the parameters $\nu$ and $\kappa$ are given. In
particular
\begin{eqnarray}
\label{u0}
U_0(t)&=&-\frac{(\nu+\lambda_3-\lambda_5)(\nu-\lambda_3-\lambda_5)}{4\,\nu^2}\,A\ t^\nu
- \frac{(\nu-\lambda_3+\lambda_5)(\nu+\lambda_3+\lambda_5)}
{4\,\nu^2}\,A^{-1}\ t^{-\nu}\nonumber\\
&+&\frac{\nu^2 -\lambda_{3}^2 +\lambda_{5}^2}{2\,\nu^2}\ .
\end{eqnarray}
where $\lambda_5 = \lambda_4-1$ and $A$ relates to $\kappa$ as
\begin{eqnarray}\label{Akappa}
A=\frac{(\nu-\lambda_3+\lambda_5)(\nu+\lambda_3+\lambda_5)}{4\, \nu^2}\ \kappa\ .
\end{eqnarray}

The expansion \eqref{doubleser} allows one to develop expansion of the
accessory parameter \eqref{cfx} in power series of $x$ as follows. For small
$x$, we are interested in the solution $y(t)$ satisfying \eqref{ydef} with
$\kappa \gg 1$; if this condition is met, $y(t)$ develops poles at
sufficiently small values of $t$, where $1/y(t)$ is still dominated by the
low-$n$ terms of the expansion \eqref{doubleser}. Let us make technical
assumption that $\Re  e(\nu)$ is small (in fact we only need $\Re e( \nu) <1$),
so that the term $U_0(t)$ in \eqref{doubleser} dominates at small $t$.
Neglecting all terms with $n>0$ leads to quadratic equation which determines
points $t=x$, $|x|\ll 1$ such that $1/y(x)=0$. The roots of this equation
\begin{eqnarray}\label{xnuA}
x^\nu = A^{-1} \qquad \text{and} \qquad x^\nu = A^{-1}\ \frac{(\nu-\lambda_3+\lambda_5)
(\nu+\lambda_3+\lambda_5)}{(\nu+\lambda_3-\lambda_5)(\nu-\lambda_3-\lambda_5)}
\end{eqnarray}
provide potential initial approximation for the position of the pole $t=x$ in
\eqref{ydef}. It is not difficult to check that only the first of these
roots corresponds to the correct sign of the residue at the pole, as show in
\eqref{ydef}; the second root would lead to the opposite sign, and thus to
the wrong value $\frac{1-(2-\lambda_4)^2}{4}$ instead of
$\frac{1-\lambda_{4}^2}{4}$ of the classical dimension $\delta_4$. With this
choice, one can systematically take into account the higher-$n$ terms in the
expansion \eqref{doubleser}, which leads to the power-like corrections to the
equation determining the pole position $x$ in terms of $\nu$ and $\kappa$,
\begin{eqnarray}
A\,x^\nu = 1 + a_1\,x + a_2\,x^2 + \ldots \ ,
\end{eqnarray}
where $A$ is given by \eqref{Akappa}, and the coefficients $a_k$ are
systematically computed order by order in terms of $A$ and $\nu$, for example,
\bea\label{sopaopsaoa}
a_1=-\frac{\nu}{2}\
\bigg(1+
\frac{ 16\, (\delta_1 - \delta_2) (\delta_3 - \delta_4) }{(1 - \nu^2)^2}\,\bigg)\ .
\eea
Expressions for the higher coefficients quickly become rather cumbersome, and we
do not present them here. Finally, according to \eqref{cxnu},
the accessory parameter $C(x,\nu)$ can be expressed through the slope and the
curvature of the function $U(t)=1/y(t)$ at its zero $t=x$,
\begin{eqnarray}
C=\frac{1}{2}\, x(1-x)\, \Big[\, U''(t)+2t\, \big(U'(t)\big)^2\Big]_{t=x}
+
\frac{\delta_3-\delta_1+\delta_2-\delta_4}{x}+
\frac{\delta_2+\delta_3+\delta_4-\delta_1}{1-x}\ .
\end{eqnarray}
Explicit calculation yields
\begin{eqnarray}
C&=&\frac{\delta_\nu-\delta_1-\delta_2}{x}+
\frac{(\delta_\nu-\delta_1+\delta_2)(\delta_\nu-\delta_4+\delta_3)}{2\,\delta_\nu}+
\bigg(\,\frac{(\delta_\nu-\delta_1+\delta_2)(\delta_\nu-\delta_4+\delta_3)}{2\,\delta_\nu}\nonumber\\
&+&\frac{\big(\,\delta_\nu^2+2\,\delta_\nu\, (\delta_1+\delta_2)-3\,(\delta_1-\delta_2)^2\,\big)\big(\,\delta^2_\nu+
2\,\delta_\nu\,(\delta_3+\delta_4)-
3\,(\delta_3-\delta_4)^2\,\big)}{32\,\delta^2_\nu\, (\delta_\nu+\frac{3}{4})}\\
&-&
\frac{\big(\,(\delta_1-\delta_2)^2-\delta^2_\nu\,\big)
\big(\,(\delta_3-\delta_4)^2-\delta^2_\nu\,\big)}
{8\,\delta_\nu^3}\,\bigg)\ x +O(x^2)\ , \nonumber
\end{eqnarray}
in complete agreement with the classical limit of well known expansion
coefficients of the four point conformal block.

\section{Remarks}

It is well known that the Painlev\'e VI equation describes the isomonodromic
deformations of the second order Fuchsian system with four singular points.
Consider the linear problem
\begin{eqnarray}\label{fuksian}
\partial_z {\boldsymbol \Psi}(z) = {\boldsymbol  A}(z)\,{\boldsymbol \Psi(z)}\,,
\qquad {\boldsymbol  A}(z) = \frac{{\boldsymbol  A}_1}{z}+\frac{{\boldsymbol A}_2}{z-t} +
\frac{{\boldsymbol A}_3}{z-1}\ ,
\end{eqnarray}
where ${\boldsymbol  A}_i$ are constant (i.e. $z$-independent) traceless $2\times 2$
matrices with ${\rm det}({\boldsymbol  A}_i) = -\frac{\lambda_{i}^2}{4}$, and ${\boldsymbol \Psi} =
\column{\psi_+}{\psi_-}$.  If
${\boldsymbol A}_4= - \sum_{i=1}^3 {\boldsymbol A}_i \neq 0$ then there are
four regular singularities $t$, $0$, $1$ and $\infty$. It is convenient to
choose the gauge such that ${\boldsymbol  A}_4 ={\rm diag}(\lambda_{4}/2,-\lambda_{4}/2)$, so that
the off-diagonal elements of the matrix ${\boldsymbol  A}(z)$ in \eqref{fuksian} decay as $1/z^2$ as
$z\to\infty$.  Define $y$ as the zero of the off-diagonal matrix element
$B(z)$ of the matrix
\begin{eqnarray}
{\boldsymbol  A}(z) =\matrix{A(z)}{\ \ \,B(z)}{C(z)}{-A(z)}
\label{eq:abcdm}
\end{eqnarray} in
\eqref{fuksian}, which then must have the form
\begin{eqnarray}
B(z) = k\ \frac{y-z}{z(z-1)(z-t)}\,.
\end{eqnarray}
The parameter $k$ can be set to $1$ by the remaining gauge transformation.

The linear Fuchsian systems admit monodromy preserving deformations, generally
expressed by the Schlesinger's equations \cite{Schlesinger,Jimbo:1981}. For
the system \eqref{fuksian},
these equations reduce to the Painlev\'e VI equation \eqref{painleve6}
for the parameter $y$  as the function of $t$ (with the equations for
the remaining elements of the matrices ${\boldsymbol   A}_i$ dependent of
this function).

The Darboux coordinates $({\nu}, {\mu})$ of the $\mathbb{SL}(2,\mathbb{C})$
flat connection ${\boldsymbol    A}(z)$ defined in   \cite{Nekrasov:2011bc}
can be computed at any point of the Schlesinger-Painlev\'e VI flow, in
particular in the limit $t \to 0$. In this limit the connection
\eqref{fuksian} reduces to a pair of hypergeometric connections on the
$3$-punctured spheres, connected by certain gauge transformation, and can be
evaluated in closed form. Thus, the coordinate ${\nu}$ is determined in this
limit from the eigenvalues of ${\boldsymbol A}_{1} +{\boldsymbol A}_{2}$,
confirming that it coincides with the parameter $\nu$ in \eqref{ydef}.
Similarly, calculating the connection at $t\to 0$ allows one to relate the
coordinate $\mu$ to the parameter $\kappa$ in \eqref{ydef}, in full agreement
with \eqref{eq:muco}, \eqref{muxnu}. We note that this agreement can be
regarded as an alternative derivation of \eqref{fact} which, unlike our
arguments in Section\,\ref{sec3}, avoids any references to quantum conformal
blocks. However, we believe that the arguments in Section\,\ref{sec3} are
more compact and suggestive.

The $n> 4$ generalization of our formalism seems to be straightforward. Let ${\CalA}_{0,n}$
be the moduli space of representations of the fundamental group of the $n$-punctured
sphere into $\mathbb{SL}(2,\mathbb{C})$ with the fixed conjugacy classes of the elementary
loops. It is endowed with the symplectic form $\Omega$ whose Darboux coordinates
$({\nu}_{i}, {\mu}_{i})$, $i = 1, \ldots  n-3$, are defined in
\cite{Nekrasov:2011bc} relative to a pant decomposition of the $n$-punctured sphere.
On the other hand, the space ${\CalX}_{n}$ of residues
${\boldsymbol A}_{i}$, $i = 1, \ldots  n$,
with fixed eigenvalues
${\rm Tr}({\boldsymbol A}_{i}^{2}) =
\frac 12\, {\lambda}_{i}^2$,  of the meromorphic ${\mathfrak sl}(2)$-connection
\begin{eqnarray}
{\boldsymbol A}(z) = \sum_{i=1}^{n}   \frac{{\boldsymbol A}_{i}}{z - z_{i}}
\label{eq:merm}
\end{eqnarray}
obeying the moment map equation
\begin{eqnarray}
\sum_{i=1}^{n} {\boldsymbol A}_{i} = 0
\label{eq:mmap}
\end{eqnarray}
and considered up to the simultaneous $\mathbb{SL}(2,\mathbb{C})$ conjugation, is a
symplectic manifold, with the Darboux coordinates $(p_i, y_i),\
i=1,\ldots n-3,$ defined relative to a point $(z_{1}, \ldots  z_{n})$
 on the moduli space ${\CalM}_{0,n}$ of $n$-punctured sphere. Let us parameterize
points of ${\CalM}_{0,n}$ by ${\bf z} =\{z_{1}, \ldots  z_{n-3}\}$, by
setting $ z_{n-2} = 0$,\ $z_{n-1} = 1$, $z_{n} = \infty$ (this is the same
convention as we adopted in Sections\,\ref{sec1} and \ref{sect2} for generic
$n$, but is slightly different from the conventions \eqref{z1..4} chosen for
$n=4$ in the main body of the paper). The coordinates $y_{i}$ are the zeros
of the matrix element $B(z)$ defined in the general case in \eqref{eq:abcdm}
in the gauge where ${\boldsymbol A}_{n}$ is diagonal. The coordinates $p_{i}$
are the eigenvalues of ${\boldsymbol A}(y_{i})$ (see \cite{sklyanin,
Sklyanin(1995), Krichever:2001cx, dubrovin} for precise definitions). The map
${\Phi}: {\CalM}_{0,n} \times {\CalX}_{n} \longrightarrow {\CalA}_{0,n}$
sends the set of residues to the monodromy data of the connection
\eqref{eq:merm}. We have:
\begin{eqnarray}
{\Phi}^{*} {\Omega} = \sum_{i=1}^{n-3} \rd p_{i} \wedge
\rd y_{i} - \sum_{i=1}^{n-3} \rd h_{i} \wedge \rd z_{i}\ ,
\label{eq:pllb}
\end{eqnarray}
where $h_{i}({\bf p},{\bf y},{\bf z})$ are the Hamiltonians of the
Schlesinger flows, generating the isomonodromic deformations of
\eqref{eq:merm}. Here and below ${\bf p}=\{p_i\}\,,\ {\bf y} = \{y_i\}$, etc.
Writing the left hand side of \eqref{eq:pllb} as $\rd {\boldsymbol\nu} \wedge
\rd {\boldsymbol\mu}$ and taking the $\rd$-antiderivative we arrive at the
equation:
\begin{eqnarray}
\rd S =
\sum_{i=1}^{n-3} \left( {\mu}_{i}\, \rd{\nu}_{i} + p_{i}\, \rd y_{i} -
h_{i}\, \rd z_{i} \right)
\label{eq:mast}
\end{eqnarray}
for some master function $S( {\bf z}, {\bf y}, {\boldsymbol \nu})$, which
solves the Hamilton-Jacobi equations
\begin{eqnarray}
h_{i} \left( \frac{{\partial}S}{{\partial}{\bf y}}, {\bf y}, {\bf z} \right)  = -
\frac{{\partial}S}{{\partial}z_{i}}\,, \qquad i = 1, \ldots  n-3
\label{eq:hamjac}
\end{eqnarray}
and can be related in a standard way to the action evaluated on a solution of
the Schlesinger equation. The function $S$ should be related to the $b \to 0$
limit of the $2n-3$-point conformal block \footnote{The relation of the
classical limit of the conformal block \eqref{n-3+n} to the Schlesinger flow
was pointed out in \cite{Teschner:2010je}.}
\begin{eqnarray}\label{n-3+n}
\Big\langle\, \prod_{i=1}^{n-3} V_{(2,1)}(y_{i}) \prod_{i=1}^{n}
V_{{\Delta}_{i}} (z_{i})\, \Big\rangle_{b\to 0}\ .
\end{eqnarray}
To make contact with the twisted superpotential $W ({\boldsymbol \nu},{\bf
q})$ of the linear quiver theory with the gauge group
$\big(\mathbb{SU}(2)\big)^{\otimes(n-3)}$
in the context of the AGT correspondence,
with the exponentiated complexified coupling $q_{i}$ of the $i^{\rm th}$
gauge factor given by $z_{i}/z_{i+1}$, proceed as follows. First pick a path
${\bf z}(t)$ in $\overline{{\CalM}_{0,n}}$ with $t \in \mathbb{CP}^{1}$,
which originates at ${\bf z}(0)=0$ in such a way that $0 < |\,z_{1} | \ll
|\,z_{2} | \ll | \,z_{3} | \ll \ldots \ll |\,z_{n-3}| \ll 1$. This
corresponds to the maximal degeneration of the $n$-punctured sphere at $t=0$,
associated with the dual diagram in Fig.\,\ref{fig1} (with the legs $1$ and
$n-2$ interchanged). Also, let ${\bf z}={\bf z}(1)$. Now, consider a solution
of Schlesinger equations restricted onto this path. The details of the path
are not essential except for the global monodromy properties, because the
Hamiltonians $h_{i}$ obey
\begin{eqnarray}
{\partial}_{z_{i}} h_{j} - {\partial}_{z_{j}} h_{i} = 0\ ,
\qquad \{ h_{i}, h_{j} \}_{\bf p, y} = 0\ .
\end{eqnarray}
One must pick the solution such that for $t=1$ the corresponding zeroes
$y_{i}$ are equal to the ``moving points'' $z_{i}$: $y_{i}({\bf z}(1)) =
z_{i}$. Then integrate $\int_{0}^{1}\, \left( {\bf p}\, {\dot {\bf y}} - {\bf
h}\, {\dot {\bf z}} \right) \, \rd t$ and subtract the logarithmically
divergent terms to arrive at $W ({\boldsymbol \nu},{\bf z})$.

Potential developments of the relation discussed in this note include its
extension to the general classical conformal blocks along the lines described
above, in particular analysis of the suitable solutions of the Schlesinger
equations, and its possible applications to the study of the analytic
properties. We hope to come back to these problems in the future.

\section*{Acknowledgments}

Part of this work was done during the visit of the  second author to IPhT at
CEA Saclay in May-June 2012. SL would like to express his sincere gratitude
to members of the laboratory and especially Didina  Serban and Ivan Kostov
for their kind hospitality and interesting discussions.

\bigskip
\noindent Research of NN was supported in part by RFBR grants 12-02-00594,
12-01-00525, by Agence Nationale de Recherche via the grant ANR 12 BS05 003
02,  by Simons Foundation, and by Stony Brook Foundation. AZ gratefully
acknowledges hospitality of the Simons Center for Geometry and Physics during
the early stages of this project. Research of AZ is supported by the DOE
under grant DE-FG02-96ER40959, and by the BSF grant.


\begin{thebibliography}{99}


\bibitem{Belavin:1984vu}
  A.~A.~Belavin, A.~M.~Polyakov and A.~B.~Zamolodchikov,
  ``Infinite conformal symmetry in two-dimensional Quantum Field Theory,''
  Nucl.\ Phys.\ B {\bf 241}, 333 (1984).


\bibitem{Nekrasov:2004sem}
N.~Nekrasov, 	
``On the BPS/CFT correspondence,'' seminar at
the string theory group, University of Amsterdam, Feb. 3, 2004.



\bibitem{Nekrasov:2002qd}
  N.~A.~Nekrasov,
  ``Seiberg-Witten prepotential from instanton counting,''
  Adv.\ Theor.\ Math.\ Phys.\  {\bf 7}, 831 (2004)
  [arXiv:hep-th/0206161].

\bibitem{Losev:2003py}
  A.~S.~Losev, A.~Marshakov and N.~A.~Nekrasov,
  ``Small instantons, little strings and free fermions,''
  In: Shifman, M. (ed.) et al.: From fields to strings, vol. 1, 581-621
  [arXiv:hep-th/0302191].


\bibitem{Nekrasov:2003rj}
  N.~Nekrasov and A.~Okounkov,
  ``Seiberg-Witten theory and random partitions''
  [arXiv:hep-th/0306238].

  \bibitem{Alday:2009aq}
  L.~F.~Alday, D.~Gaiotto and Y.~Tachikawa,
  ``Liouville correlation functions from four-dimensional gauge theories,''
  Lett.\ Math.\ Phys.\  {\bf 91}, 167 (2010)
  [arXiv:hep-th/0906.3219].

\bibitem{Zamolodchikov:1987}
  Al.~B.~Zamolodchikov,
  ``Conformal symmetry in two-dimensions: recursion representation of conformal block,''
  Teor.\ Mat.\ Fiz.\  {\bf 73}, No.1, 103 (1987)
(Theor. ~Math. ~Phys., {\bf 53}, 1088 (1987)).


\bibitem{Zamolodchikov:1995aa}
  A.~B.~Zamolodchikov and Al.~B.~Zamolodchikov,
  ``Structure constants and conformal bootstrap in Liouville field theory,''
  Nucl.\ Phys.\  B {\bf 477}, 577 (1996)
  [arXiv:hep-th/9506136].


\bibitem{Gamayun:2012ma}
  O.~Gamayun, N.~Iorgov and O.~Lisovyy,
  ``Conformal field theory of Painlev\'e VI,''
  JHEP {\bf 1210}, 038 (2012)
  [Erratum-ibid.\  {\bf 1210}, 183 (2012)]
  [arXiv:hep-th/1207.0787].


\bibitem{Zamolodchikov:1985ie}
  Al.~B.~Zamolodchikov,
  ``Conformal symmetry in two-dimensions: an explicit recurrence formula for the conformal partial wave amplitude,''
  Commun.\ Math.\ Phys.\  {\bf 96}, 419 (1984).

\bibitem{Alba:2011} V. A. Alba, V. A.
    Fateev, A. V. Litvinov and G. M. Tarnopolsky, ``On combinatorial expansion
of the conformal blocks
    arising from AGT conjecture,'' Lett.~Math.~Phys. {\bf 98}, 33 (2011)
    [arXiv:hep-th/1012.1312].


\bibitem{Nekrasov:2009rc}
  N.~A.~Nekrasov and S.~L.~Shatashvili,
  ``Quantization of integrable systems and four dimensional gauge theories''
  [arXiv:hep-th/0908.4052].

\bibitem{Nekrasov:2013}
  N.~Nekrasov, V.~Pestun and S.~Shatashvili,
  ``Quantum geometry and ${\CalN}=2$ quiver gauge theories,''
  to appear.

\bibitem{ab} M.~F.~Atiyah and R.~Bott,
``The Yang-Mills equations over Riemann surfaces,''
Phil. Trans. R. Soc. Lond.  A {\bf 308},  523-615 (1982).



\bibitem{Bilal:1990wn}
  A.~Bilal, V.~V.~Fock and I.~I.~Kogan,
  ``On the origin of W algebras,''
  Nucl.\ Phys.\ B {\bf 359}, 635 (1991).


\bibitem{boalch} P.~Boalch,
``Symplectic manifolds and isomonodromic deformations,''
Adv. Math. {\bf 163},  137-205  (2001).

\bibitem{Nekrasov:2011bc}
  N.~Nekrasov, A.~Rosly and S.~Shatashvili,
  ``Darboux coordinates, Yang-Yang functional, and gauge theory,''
  Nucl.\ Phys.\ Proc.\ Suppl.\  {\bf 216}, 69 (2011)
  [arXiv:hep-th/1103.3919].




\bibitem{zt}
L.~Takhtajan and P.~Zograf,
``Action for the Liouville equation as a generating
 function for the accessory
parameters and as a potential for the Weil-Petersson metric on the
Teichm${\ddot{\rm u}}$ller space,''
Funkts. Anal. Prilozh. {\bf 19}, 67 (1985)
[Engl. transl.: Funct. Anal. Appl. {\bf 19}, 219 (1986)];
``On the Liouville equation, accessory parameters and the geometry of the
Teichm${\ddot{\rm u}}$ller space for the Riemann surfaces of genus $0$,''
Mat. Sbornik {\bf 132}, 147 (1987) [Engl. transl.: Math. USSR Sbornik {\bf 60}, 143  (1988)].

\bibitem{klein}
F.~Klein,
``Neue Beitr${\ddot{\rm a}}$ge zur
Riemann'schen Functionentheorie,'' Math. Ann. {\bf 21}, 201 (1883).

\bibitem{poincare}
A.~Poincare,
``Sur les groupes des equations lineaires'', Acta Math., {\bf 4}, 201 (1884);\\
``Les fonctions fuchsiennes et l'equation $\triangle u=\exp(u)$,''
J. Math. Pures Appl., {\bf 5}, 157, (1898).



\bibitem{Teschner:2010je}
 J.~Teschner,
``Quantization of the Hitchin moduli spaces, Liouville theory,
and the geometric Langlands correspondence I,''
  Adv.\ Theor.\ Math.\ Phys.\  {\bf 15}, 471 (2011)
  [arXiv:hep-th/1005.2846].

\bibitem{Lukyanov2011} S. L. Lukyanov, ``Critical values of the Yang-Yang
    functional in the quantum sine-Gordon model,'' Nucl.Phys. {B} {\bf 853},
    475-507 (2011)  [arXiv/hep-th:1105.2836].


\bibitem{Guzzetti2011}
D.~Guzzetti,
``Tabulation of PVI transcendents and parametrization formulas,''
Nonlinearity {\bf 25}, 3235 (2012)
[arXiv:math.CA/1108.3401].

\bibitem{Schlesinger}
L.~Schlesinger,
``Ueber eine klasse von differentsial system beliebliger ordnung mit
festen kritischer punkten'',
J. fur  Math. {\bf 141}, 96 (1912).

\bibitem{Jimbo:1981} M.~Jimbo, T.~Miwa and  K.~Ueno,
``Monodromy preserving
deformations of linear differential equations with rational coefficients (1),''
  Physica D {\bf 2}, 407 (1981).



\bibitem{sklyanin} E.~K.~Sklyanin,
``Goryachev-Chaplygin top
and the inverse scattering method,''
Zapiski Nauch. Semin. LOMI133, 236¿257 (1984)
[Engl. transl.: J. Soviet Math. {\bf 31}, 3417¿3431 (1985)];

\bibitem{Sklyanin(1995)} E.~K.~Sklyanin,
``Separation of variables -- new trends,'' Prog. Theor. Phys. Suppl. {\bf 118}, 35-60 (1995)
[arXiv:solv-int/9504001].

\bibitem{Krichever:2001cx}
  I.~Krichever,
  ``Isomonodromy equations on algebraic curves, canonical
transformations and Whitham equations''
  [arXiv:hep-th/0112096].

\bibitem{dubrovin} B.~Dubrovin,
M.~Mazzocco, ``Canonical structure and symmetries of
the Schlesinger equations'' [arXiv:math/0311261].


\bibitem{kashani-poor} A.-K.~Kashani-Poor, J.~Troost, 
``Transformations of
    spherical blocks'' [arXiv:hep-th/1305.7408].

\end{thebibliography}
\end{document}